\documentclass[preprint]{aastex}
\usepackage{color}

\begin{document}



\title{The Orbit of the Companion to HD 100453A:  Binary-Driven Spiral Arms in a Protoplanetary Disk}
\color{black}
\shorttitle{Scorpion-1b MagAO}
\shortauthors{placeholder}
\author{Kevin Wagner\altaffilmark{1,7,8}$^{\star}$, Ruobing Dong\altaffilmark{1,9}, Patrick Sheehan\altaffilmark{2}, D\'aniel Apai\altaffilmark{1,3,8}, Markus Kasper\altaffilmark{4}, Melissa McClure\altaffilmark{4}, Katie M. Morzinski\altaffilmark{1}, Laird Close\altaffilmark{1}, Jared Males\altaffilmark{1}, Phil Hinz\altaffilmark{1}, Sascha P. Quanz\altaffilmark{5}, and Jeffrey Fung\altaffilmark{6}}
%


\altaffiltext{1}{Steward Observatory, University of Arizona, 933 North Cherry Avenue, Tucson, AZ 85721}

\altaffiltext{2}{Homer L. Dodge Department of Physics and Astronomy, University of Oklahoma, Norman, OK 73019}
\altaffiltext{3}{Lunar and Planetary Laboratory, University of Arizona, 1629 E University Blvd, Tucson, AZ 85721}
\altaffiltext{4}{European Southern Observatory, Karl-–Schwarzschild-–Str. 2, D–85748 Garching, Germany}
\altaffiltext{5}{Institute for Astronomy, ETH Zurich, Wolfgang-Pauli-Strasse 27, 8093 Zurich, Switzerland}
\altaffiltext{6}{Department of Astronomy, University of California at Berkeley, Berkeley, CA 94720}

\altaffiltext{7}{National Science Foundation Graduate Research Fellow}
\altaffiltext{8}{NASA NExSS \textit{Earths in Other Solar Systems} Team}
\altaffiltext{9}{Bart J. Bok Postdoctoral Fellow}

\altaffiltext{$\star$}{Correspondence to: kwagner@as.arizona.edu}

\begin{abstract}

HD 100453AB is a 10$\pm$2 Myr old binary whose protoplanetary disk was recently revealed to host a global two-armed spiral structure. Given the relatively small projected separation of the binary (1$\farcs$05, or $\sim$108 au), gravitational perturbations by the binary seemed to be a likely driving force behind the formation of the spiral arms. However, the orbit of these stars remained poorly understood, which prevented a proper treatment of the dynamical influence of the companion on the disk. We observed HD~100453AB between 2015-2017 utilizing extreme adaptive optics systems on the \textit{Very Large Telescope} and \textit{Magellan Clay Telescope}. We combined the astrometry from these observations with published data to constrain the parameters of the binary's orbit to \textit{a}=1$\farcs$06$\pm0\farcs09$, \textit{e}$=0.17\pm0.07$, and \textit{i}=$32.5^{\circ}\pm6.5^\circ$. We utilized publicly available ALMA $^{12}$CO data to constrain the inclination of the disk, $i_{disk}\sim $ 28$^{\circ}$, which is relatively co-planar with the orbit of the companion and consistent with previous estimates from scattered light images. Finally, we input these constraints into hydrodynamic and radiative transfer simulations to model the structural evolution of the disk. We find that the spiral structure and truncation of the circumprimary disk in HD 100453 are consistent with a companion-driven origin. Furthermore, we find that the primary star's rotation, its outer disk, and the companion exhibit roughly the same direction of angular momentum, and thus the system likely formed from the same parent body of material.

\end{abstract}


Submitted to the \textit{Astrophysical Journal} on Oct 28, 2017; accepted Jan 11, 2018.

\section{Introduction}

Structures in protoplanetary disks are often used to predict the possible properties of planets that may be responsible for their origin. Once a planetary core has gained sufficient mass to undergo run-away gas accretion, it may interact strongly with the protoplanetary disk, driving large-scale disk structures, such as gaps (e.g., \citealt{Jang-Condell2012}, \citealt{Jang-Condell2013}, and \citealt{Dong2015a}), vortices (e.g., \citealt{Hammer2017}), and spiral density waves (e.g., \Citealt{Zhu2015}, \citealt{Dong2015b}). These interactions contribute to the disk evolution and may significantly alter the environment of planet formation. Recently, near-infrared (NIR) scattered light imaging has revealed such structures in a number of disks in unprecedented detail (e.g., \citealt{Muto2012}, \citealt{Grady2013}, \citealt{Benisty2015}, \citealt{Wagner2015}, \citealt{Akiyama2016}). However, the hypothesized companions responsible for these disk features are rarely seen themselves, leading to ambiguity in the origin of the structures.

In this work we focus on ``grand design" spiral protoplanetary disks, those with a global two-armed spiral structure, which have recently been detected in a few protoplanetary disks (SAO 206462: \citealt{Muto2012}, MWC~758: \citealt{Grady2013}, and HD~100453: \citealt{Wagner2015}). Currently there exist more alternate hypotheses for the generation of these features than there exists sources to test their predictions. Most models include a massive perturbing companion (either a planet or a low mass star, e.g., \citealt{Dong2015b} and \citealt{Zhu2015}), though notably some do not require the presence of any companion (e.g. \citealt{Montesinos2016}, \citealt{Kama2016}, \citealt{Benisty2017}). 

In \cite{Wagner2015}, we presented the discovery of a two-armed spiral structure in the disk of the 10$\pm$2 Myr old Herbig Ae star HD 100453A (d=103$\pm$3 pc, \citealt{GAIA1}; $M_{\star}$=1.7M$_{\odot}$, \citealt{Dominik2003}). The primary A-star hosts an M-dwarf companion with a mass of $\sim$0.2$M_{\odot}$ and an angular separation of 1$\farcs$05, corresponding to a projected physical separation of $\sim$108 au if the orbit is seen close to face-on (\citealt{Chen2006}, \citealt{Collins2009}, and Figure 1). Assuming a near face-on and circular orbit of HD 100453AB, \cite{Dong2016a} showed that the spiral structures are reproduced in three-dimensional hydrodynamic and radiative transfer simulations with a remarkable resemblance to the scattered light images of the disk. 

\begin{figure}[htpb]
\figurenum{1}
\epsscale{0.7}
\plotone{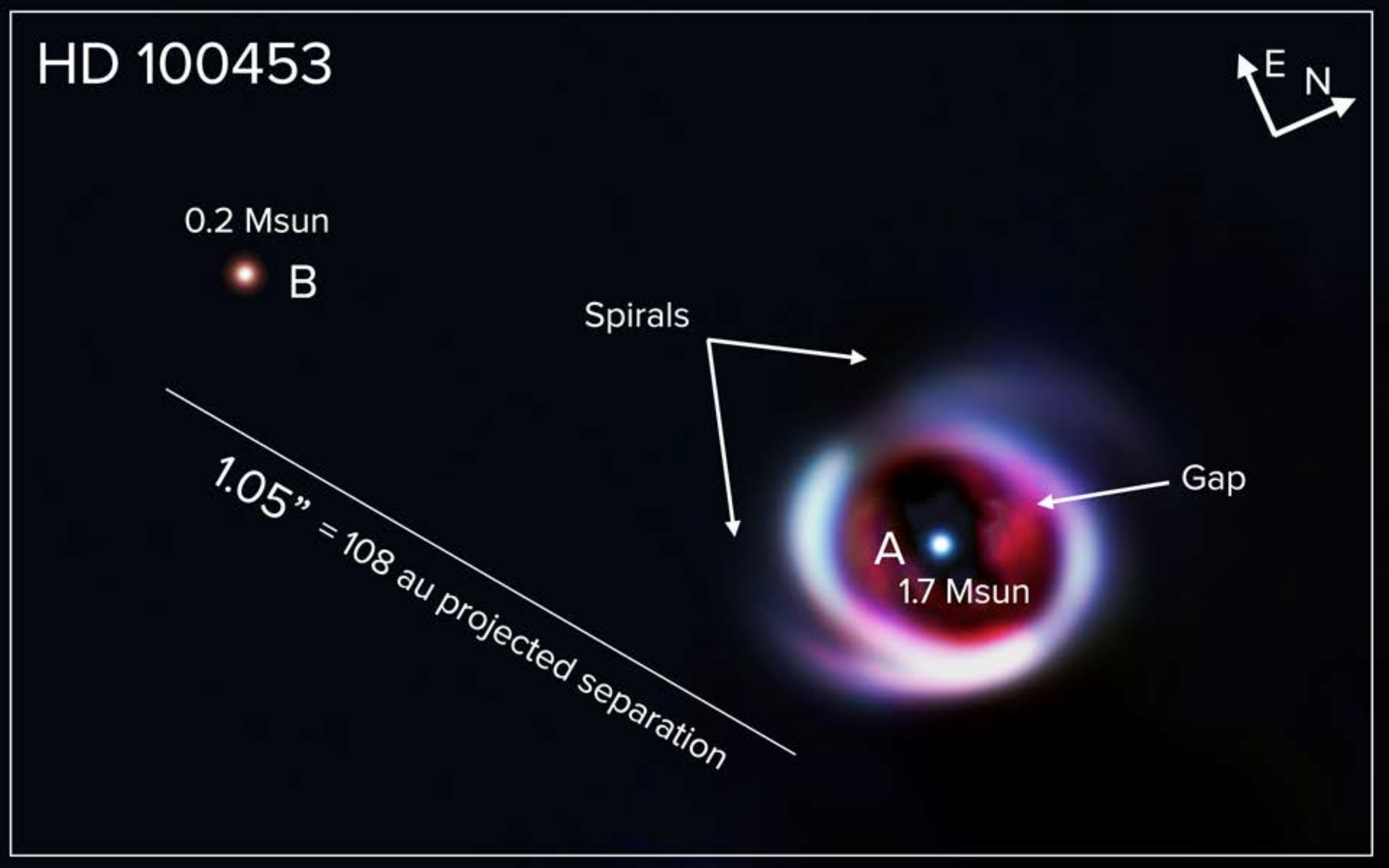}
\caption{\footnotesize Schematic diagram of the HD 100453 system constructed from the SPHERE data in \cite{Wagner2015}. The primary A-star hosts an M-dwarf companion and a spiral protoplanetary disk. The disk also hosts a gap comparable in size to Uranus's orbit about the Sun ($R\sim$20 au). The actual details inside of the gap are obscured by the coronagraph and subtraction residuals, and should not be physically interpreted from this image. The brightness of the stars with respect to the disk and to each other have been reduced for clarity, and the image colors of red, green, blue are composed of $Y-$, $J-$, and $H$-bands, respectively.}
\end{figure}

To our knowledge, no effort has been undertaken to fit the orbit of the binary, which is necessary to accurately model the effects of HD 100453B on the disk, and to test whether this body may be responsible for the spiral structure. Indeed, \cite{Montesinos2016} provide an alternate mechanism for generating the spiral structure in this disk (further discussed in \citealt{Benisty2017}) without invoking the gravitational perturbations from the companion, which may be the case if it is in fact on a much wider orbit. Thus, knowledge of the binary's true orbit may be used to discriminate between these hypotheses for the generation of the spiral arms, thereby providing constraints on the dynamics and evolution of this disk and, potentially, provide more general insights into disks with two-armed spiral structures. If the arms are induced by the companion, then any other effects, such as the influence of potentially embedded planets, disk photoevaporation, gravitational instability, or shadow-induced density variations, are not necessary to explain the origin of the spiral arms, and furthermore must not be strong enough to significantly alter their morphology. 


In this paper we present a fourteen-year baseline of astrometric measurements from adaptive optics imaging with the \textit{Very Large Telescope} Nasmyth Adaptive Optics System and Near-Infrared Imager (VLT/NACO), Spectro-Polarimetric High Contrast Exoplanet Research instrument (VLT/SPHERE), and \textit{Magellan} Adaptive Optics (MagAO) system. We use these data to provide the first orbital analysis of the HD 100453AB system. We also present disk kinematic modelling of publicly available \textit{Atacama Large Millimeter/Sub-millimeter Array} (ALMA) CO data to measure the inclination of the disk around HD 100453A, and combine these in our analysis with other disk inclination values from the literature. We explore the resultant disk structure through hydrodynamic and radiative transfer modelling. Finally, we assemble the known information on the system to provide the first complete picture of the geometry in the HD 100453 system, and discuss how this fits into the framework of the system's evolution. The datasets used in this work and data reductions are described in $\S$2. The disk and orbital modelling methodologies are described in $\S$3. The results are shown in $\S$4, and our interpretation is discussed in $\S$5. Finally, we include a brief summary and concluding remarks in $\S$6. 

\section{Observations and Data Reduction}

\subsection{VLT/SPHERE Observations}

We observed HD 100453 with VLT/SPHERE \citep{Beuzit2008} under program ID: 095.C-0389 (PI: Apai). We obtained data on three separate nights during 2015 and 2016 in IRDIFS and IRDIFS-Extended modes, providing simultaneous dual-band imaging with the IRDIS camera \citep{Vigan2010} and integral field spectroscopy (IFS, \citealt{Claudi2008}). HD 100453B is present only in the IRDIS images due to the limited field of view of the IFS. The companion is easily seen in the individual (8 second) SPHERE exposures, though each observing sequence was typically longer (about 30 minutes) to enable imaging of the disk structure closer to the primary star (presented in \citealt{Wagner2015}). 

We processed the data using the SPHERE data reduction pipeline \citep{Pavlov2008} to correct bad pixels, subtract dark current, and divide by the instrument flat field image. The remainder of our data reduction was done using custom IDL scripts, building upon those described in \cite{Wagner2015}, \cite{Apai2016}, and \cite{Wagner2016}. We corrected the images for 0.6\% anamorphic distortion by up-scaling along the vertical of the detector \citep{Maire2016}. We then computed the star center (obscurred by the coronagraph) utilizing the position of the satellite spots generated by applying a sinusoidal pattern to SPHERE's deformable mirror in the first few frames. The centers of the four satellite spots were determined by fitting a two-dimensional Gaussian to each spot, and the intersection of the lines connecting centers of opposite spots was calculated and used as the star center. The majority of the frames (those without satellite spots) were aligned via cross-correlation and bi-linear interpolation for the sub-pixel shifts. The frames were rotated to share a common orientation of North up and East to the left by applying a clock-wise rotation of 1.75$^{\circ}$ along with a frame-by-frame determined correction (typically $\leq$0.1$^{\circ}$) due to the internal time synchronization error between the telescope and SPHERE's derotator, as described in \citep{Maire2016}. The plate scale of 12.255$\pm$0.009 mas/pixel was also obtained from \citep{Maire2016}. Finally, the images were median-combined, and then $H2$ and $H3$ ($K1$ and $K2$) were summed to form the $H23$  ($K12$) images shown in Figure 2. These observations resulted in the detection of the companion at a signal to noise ratio (SNR) of $\sim$130 in $H23$ (2016-01-21) and $\sim$100 in each $K12$ epoch (2015-04-10 and 2016-01-23), where the noise is estimated as the mean of the absolute values of flux measurements in all non-overlapping apertures at the radius of HD 100453B.

\begin{figure}[htpb]
\figurenum{2}
\epsscale{1.}
\plotone{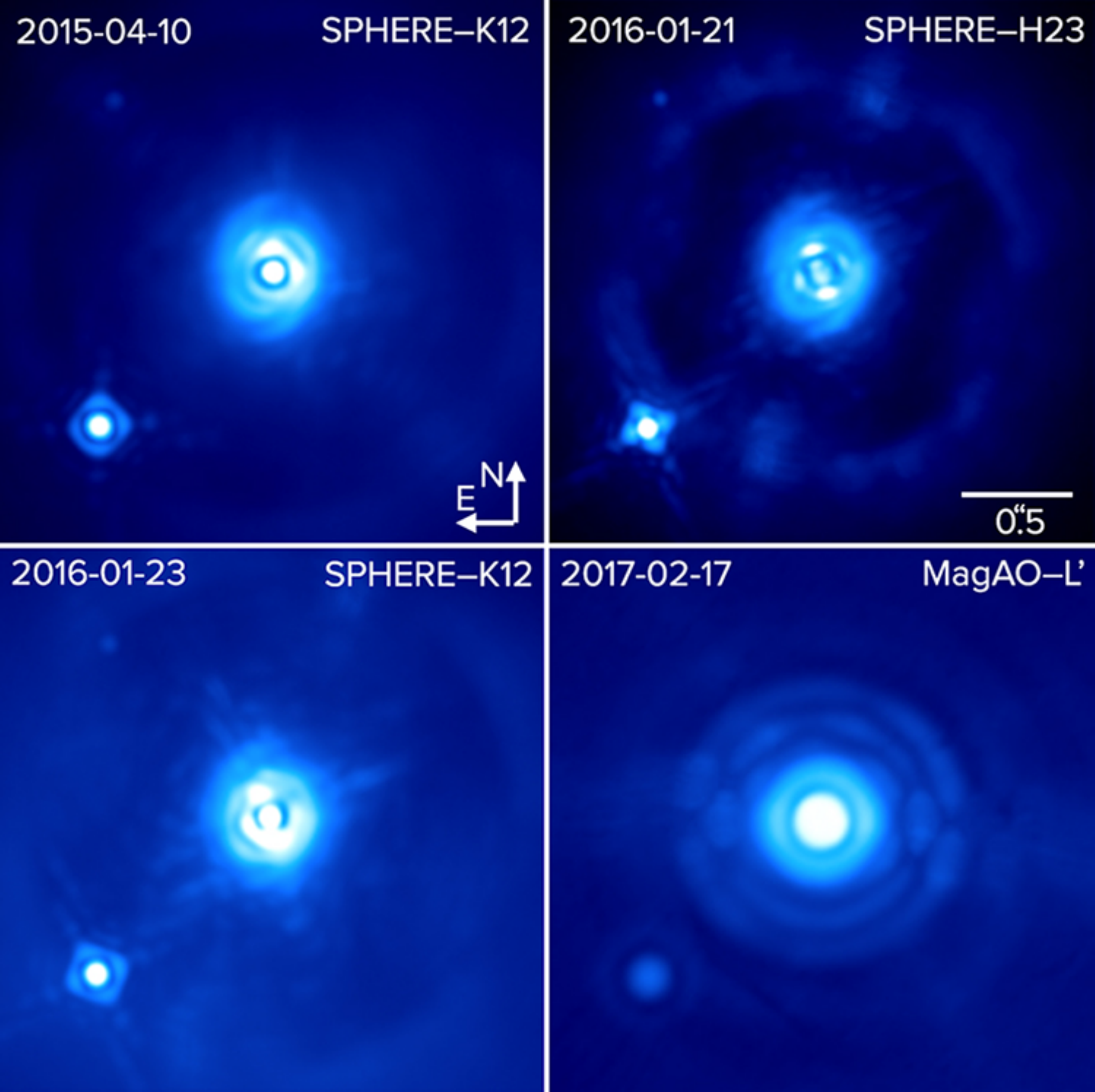}
\caption{Images of HD 100453 taken with SPHERE/IRDIS and MagAO/Clio2. The images are adjusted to a common plate scale and orientation, while the image stretch is arbitrarily adjusted in each panel to reveal the position of HD 100453B. Each of the SPHERE observations utilized a coronagraph, while the MagAO data were taken in direct mode (without a coronagraph). HD 100453B is clearly detected in each image, as well as the spiral disk in each of the SPHERE images. No effort has been made to subtract the PSF of HD 100453A from these images. }
\end{figure}

\subsection{Magellan/MagAO Observations}

We observed HD 100453 on February 17, 2017 using the Magellan Adaptive Optics System \citep{Close2012a} on the 6.5-meter \textit{Magellan Clay Telescope} at Las Campanas Observatory, Chile. We used the Clio-2 camera (\citealt{Morzinski2013}) operating in the $L^{\prime}$ (3.76 $\mu$m) filter. We obtained a total exposure time of 92 minutes with Clio-2  and 54$^\circ$ of field rotation. The airmass varied from 1.12 to 1.48 and the seeing was excellent with an average of $\sim$0$\farcs$5. We utilized a ten arcsecond nodding sequence in RA every four minutes to maintain consistent sky and source coverage. Detector integration times were kept short (0.5 sec) to keep the counts linear everywhere but the saturated PSF core. This strategy keeps the core of HD 100453B within the detector's linear response regime, allowing us to use its PSF as a photometric calibration. We corrected the images for distortion by utilizing the solution given in \cite{Morzinski2015} and verified the accuracy of the correction by applying the same step to observations of an astrometric field observed in the same night (Theta Ori B) compared to previous observations with LBT/PISCES \citep{Close2012b}. We also used these same data to determine the plate scale and instrument true North orientation post-distortion correction (15.85$\pm$1.20 mas/pixel and $2.11^\circ \pm0.16^\circ$ counter-clockwise, respectively). We processed the data using custom IDL routines to correct bad pixels, subtract the thermal background (estimated as the mean of the detector background levels before and after each nod sequence), and then aligned the images via cross-correlation and a rotational-based centering algorithm.\footnote{The centering procedure is similar to that used in \citealt{Morzinski2015}, which utilizes the rotational symmetry of the PSF along with differential imaging to find its precise center. } The images were rotated to a common orientation of North up and East left, and median-combined to create the image in the bottom right panel of Figure 2. 

\subsection{Published VLT/NACO Data}

We utilize two epochs of previously published data in our astrometric dataset. The first measurement was obtained with VLT/NACO in 2003 under program ID: 071.C-0507 (PI: Mouillet) and presented in \cite{Chen2006}. These data were reprocessed, along with an additional epoch from the same telescope/instrument in 2006 (program ID: 077.C-0570, PI: van Boekel) and presented in \cite{Collins2009}. We utilize these published separation and position angle measurements in our astrometric catalog (see \S5.1) and refer interested readers to these respective publications for the details of the data acquisition and processing. 

Since explicit detail on the astrometric calibrations of the NACO data are not included in the publications, we obtained additional information from Collins et al. through private communication. The NACO observations were obtained in field-tracking mode, and the individual image exposure times were kept short (0.35s) to limit saturation of the primary star. The positions of both HD 100453A and B were obtained by fitting Gaussian functions to their PSFs in the narrow-band Br$\gamma$ images. A true North orientation of 0$^\circ$ (up on the detector) and a platescale of 27.15 mas/pixel were obtained from the image headers.

The true North orientation\footnote{Defined such that counter-clockwise in the E-left N-up plane is the positive rotation direction.} and platescale of the detector are measured to be stable at -0.04$\pm$0.14$^{\circ}$ and 27.012$\pm$0.004 mas/pixel, respectively, throughout the period of observations (\citealt{Chauvin2010}). These are slightly different than the values assumed in \cite{Collins2009}. While these are minor differences, to maintain the highest possible degree of astrometric accuracy we have adjusted the values used in our study accordingly with the aforementioned calibrations. This results in a $\sim$1$\sigma$ change for both separation measurements (-6 mas and -5 mas for the 2003 and 2006 measurements, respectively), and a negligible $\sim$0.1$\sigma$ change in position angle measurements. We have also adopted a slightly larger measurement uncertainty after combining the uncertainties listed in \cite{Collins2009} and those on the platescale and true North orientation by summing uncertainties in quadrature.

\subsection{Astrometric Error Budget}

Since only a small fraction of the orbit of HD 100453AB has been observed, the precision in our orbit fitting is heavily reliant on the precision of our astrometric measurements. In this section we describe the uncertainties and error budget of the new astrometric measurements used in the proceeding work, which are summarized in Table 1. The primary components that enter into the final (combined) uncertainty are 1) uncertainty in the position of the primary star that is either saturated or obscurred by a coronagraph; 2) uncertainty in the measurement of the position of HD 100453B; and 3) uncertainty in the plate scale and orientation of the detector. The location of the primary star is obtained via satellite spot centering (for SPHERE data) and via rotational centering (for MagAO), described in \S 2.1 and \S 2.2, respectively, which are typically accurate to $\lesssim$0.25 pixels (\citealt{Mesa2015}, \citealt{Morzinski2015}). For both datasets, the position of HD 100453B is measured by fitting a 2D Gaussian to the source. The measurement uncertainties (1$\sigma$) are estimated as the FWHM divided by the SNR. Both instrumental PSFs are super-Nyquist sampled (3$-$4.5 pixel FWHM for SPHERE $H$ and $K$, and 7.5 pixels for MagAO $L^{\prime}$), and the source is detected at relatively high SNR in each dataset (SNR $>$ 100 for SPHERE, and $>$ 30 for MagAO), allowing for sub-pixel accuracy in the measurement of the position of HD 100453B. 
\begin{deluxetable}{cccccc}
\tabletypesize{\scriptsize}
\tablecaption{Astrometric Measurements and Error Budget}
\tablewidth{0pt}
\tablehead{
\colhead{Date} & \colhead{Separation \&} & \colhead{HD 100453B} & \colhead{HD 100453A} & \colhead{Plate Scale \&}  & \colhead{Combined} \\
\colhead{Facility-Filter} & \colhead{Position Angle} & \colhead{Uncertainty} & \colhead{Uncertainty} & \colhead{Field Orient. Unc.} &\colhead{Uncertainty} }
\startdata\




\\
2003-06-02 & 1$\farcs$049 & \multicolumn{2}{c}{0$\farcs$007}  & 0.004 mas & 0$\farcs$0070 \\
NACO-Br$\gamma$ & 127.23$^{\circ}$ &  \multicolumn{2}{c}{0.30$^\circ$}  &0.14$^{\circ}$& 0.33$^{\circ}$\\
\\
2006-06-22 & 1$\farcs$042 & \multicolumn{2}{c}{0$\farcs$005}  & 0.004 mas & 0$\farcs$0050 \\
NACO-Br$\gamma$  & 128.26$^{\circ}$ &  \multicolumn{2}{c}{0.27$^\circ$} &0.14$^{\circ}$& 0.30$^{\circ}$\\
\\
2015-04-10 & 1$\farcs$047 & 0$\farcs$00057 & 0$\farcs$003  & 0.09 mas & 0$\farcs$0031 \\
SPHERE-K12 & 131.63$^{\circ}$ & 0.031$^{\circ}$ & 0.17$^{\circ}$  &0.08$^{\circ}$& 0.19$^{\circ}$\\
\\
2016-01-21 & 1$\farcs$056 & 0$\farcs$00034 & 0$\farcs$003  & 0.09 mas & 0$\farcs$0030 \\
SPHERE-H23 & 131.95$^{\circ}$ & 0.018$^{\circ}$ & 0.17$^{\circ}$  &0.08$^{\circ}$& 0.19$^{\circ}$\\
\\
2016-01-23 & 1$\farcs$053 & 0$\farcs$00054 & 0$\farcs$003 & 0.09 mas & 0$\farcs$0030 \\
SPHERE-K12 & 131.95$^{\circ}$ & 0.029$^{\circ}$ & 0.17$^{\circ}$ &0.08$^{\circ}$& 0.19$^{\circ}$\\
\\
2017-02-17 & 1$\farcs$056 & 0$\farcs$0032 & 0$\farcs$004 & 1.8 mas & 0$\farcs$0054 \\
MagAO-L$^{\prime}$ & 132.32$^{\circ}$ & 0.18$^{\circ}$ & 0.22$^{\circ}$& 0.34$^{\circ}$ & 0.44$^{\circ}$\\

\enddata
\tablecomments{Table 1: Astrometric uncertainties for observations of HD 100453AB. The SPHERE plate scale and field orientation calibrations were obtained from \cite{Maire2016}. The MagAO plate scale and field orientation uncertainty represents the combined uncertainty from our calibrations and also the uncertainty in the fiducial plate scale and field rotation calibrations utilized in \cite{Close2012b}. The NACO data were obtained from \cite{Collins2009} and adjusted according to the platescale and true North orientation of \cite{Chauvin2010}. Since uncertainties on the positions of the two stars in \cite{Collins2009} are already combined into separation and position angle, only one value is listed for their combination. The individual uncertainties (reported here as 1$\sigma$) were combined in quadrature for the final combined uncertainties.}

\end{deluxetable}

\subsection{ALMA $^{12}$CO Observations}

HD 100453 was observed on 2016 April 23 with ALMA Band 6 during Cycle 3 (Program Code: 2015.1.00192.S, PI: van der Plas).\footnote{These data are part of an on-going campaign, and the higher resolution component of the dataset will be presented in an upcoming paper (van der Plas et al, in prep.).} During the observations forty-two 12 meter antennas were available, with baselines ranging from 12--460 meters. The Band 6 receiver was configured for $^{12}$CO 2--1 observations, with a baseband of 1,920 channels centered at 230.529 GHz with 61 kHz channel widths (0.16~km/s~velocity resolution; Hanning smoothed). J1107-4449 was used as the bandpass and flux calibrator, and J1132-5606 was used as the the gain calibrator. The on-source time was $\sim$13 minutes. 

\begin{figure}[htpb]
\figurenum{3}
\epsscale{1.0}
\plotone{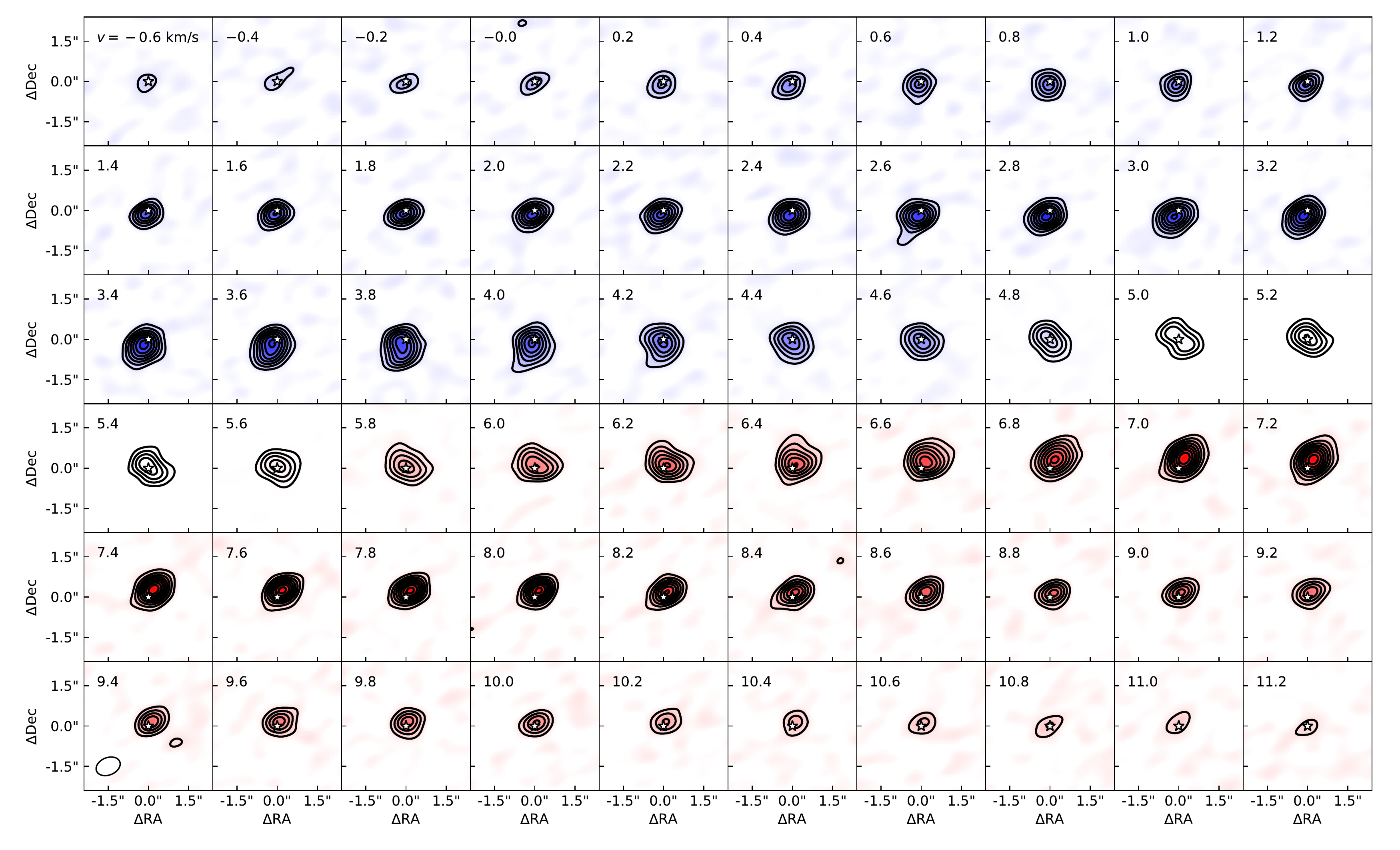}
\caption{Channel maps of $^{12}$CO 2--1 emission in the disk surrounding HD 100453A from ALMA observations taken on 2016 April 23. The contours represent integer multiples of the $3\sigma$ detection limit ($\sigma$=8.1 mJy/beam).}
\end{figure}

The data were calibrated using the ALMA pipeline in the \texttt{CASA} package. Following calibration, we imaged the $^{12}$CO 2--1 data using the CLEAN routine. Imaging was done in 0.2~km/s channels using natural weighting to enhance sensitivity in the channel maps. The CO channel maps (Figure 3) has a beam size of $0\farcs94 \times 0\farcs65$ with PA of $-67.8^{\circ}$, and a mean rms of 8.1 mJy beam$^{-1}$ in signal-free channels. The data show a clear red- and blue-shift pattern for the NW and SE sides of the disk, respectively, which is characteristic of a moderately inclined disk. Combined with the images in \cite{Benisty2017} showing scattered light to the SW from the bottom side of the disk, we can infer a counter-clockwise rotation of the disk in the plane of the sky. We note that the disk and binary are both orbiting in the same direction. In \S3.2 and \S4.3.2 we present a model representative of these data to investigate the disk kinematics and physical properties.

\section{Modelling Methodologies}

\subsection{Orbital Modelling}

We employ the grid-search and bootstrapping method of orbit-fitting presented in \cite{Wagner2016}. This technique uses the method in \cite{Meeus1998} to calculate the position angle (PA) and separation of a binary system as a function of time and the seven Keplerian orbital parameters. We constructed a grid of models and performed a chi-squared minimization to find the best-fit model, while keeping the mass of the system fixed at 1.9 M$_{\odot}$. Each trial performed seven iterations of parameter searches for a total of of 46,656 combinations of the six orbital parameters (with period and semi-major axis linked through mass). At each iteration the grid range and spacing were reduced by 50\%. The distribution of orbital parameters shown in \S 4.2 were assembled by repeating the procedure for 25,265 trials, during each of which the data points were modulated by adding Gaussian noise drawn from a distribution with mean of zero and standard deviation of unity, multiplied by the 1$\sigma$ measurement uncertainties.

\subsection{CO 2--1 Modelling}

We follow the modelling procedure outlined in \cite{Wu2017} to fit our ALMA channel maps with synthetic channel maps produced from radiative transfer models, and thereby constrain the bulk disk structure. We describe the procedure briefly here, but additional details can be found in \cite{Wu2017}. We construct a simple model of the disk, aimed at constraining the inclination of the bulk of the gas in the spiral-arm hosting outer disk. We assume that the $^{12}$CO (2--1) emission comes from a flared accretion disk that is in vertical hydrostatic equilibrium. We use a power law in radius to parameterize the temperature profile of the disk, and assume that the disk is vertically isothermal. Finally, we assume that rotation in the disk is Keplerian and that the disk is in the same plane throughout. While there is a known warp in the inner dust disk (\citealt{Benisty2017}, \citealt{Long2017}), the CO data are not of sufficient resolution to reveal whether the gas follows this same structure. Thus, our simple model mostly reflects the conditions of the outer disk, which contains the majority of disk surface area and (presumably) CO emission. Under these assumptions, the equations that govern the disk structure are:
\begin{equation}
\rho(R,z) = \frac{\Sigma(R)}{\sqrt{2\pi} \, h(R)} \, \exp\left[-\frac{1}{2}\left(\frac{z}{h(R)}\right)^2\right],
\end{equation}
\begin{equation}
\Sigma = \Sigma_0 \, \left(\frac{R}{r_c}\right)^{-\gamma} \, \exp\left[-\left(\frac{R}{r_c}\right)^{2-\gamma}\right]
\end{equation}
\begin{equation}
N_{\rm CO}(R) = \frac{X_{\rm CO} \, \Sigma(R)}{\mu \, m_{\rm H}},
\end{equation}
\begin{equation}
T(R) = T_0 \, \left(\frac{R}{1~{\rm AU}}\right)^{-q},
\end{equation}
\begin{equation}
h(R) = \left( \frac{k_b \, R^3 \, T(R)}{G \, M_* \, \mu \, m_{\rm H}} \right)^{1/2},
\end{equation}
\begin{equation}
v_k = \sqrt{\frac{G \, M_*}{r}}.
\end{equation}
where $R$ and $z$ are defined in cylindrical coordinates, $\Sigma(R)$ and $h(R)$ are the surface density and disk scale height, respectively. Here we fix the stellar mass to a value of $M_* = 1.7$ $M_{\odot}$ and the distance to a value of $d = 103$ pc based on previous estimates of HD 100453's mass \citep{Dominik2003} and distance \citep{GAIA1}. $X_{\rm CO} = 1 \times 10^{-4}$ is the CO mass abundance fraction, and $\mu = 2.37$ is the mean molecular weight. The disk is truncated at an inner radius of 0.1 AU. We also include microturbulent line broadening, which we assume is uniform throughout the disk, with a value of $\xi$ in units of km/s. Finally, we allow the star to have a systemic velocity, $v_{sys}$, which Doppler shifts the velocity center away from zero. In all, the density, temperature, and velocity structure of the system are described by the following parameters: $M_{\rm disk}$, $r_{\rm c}$, $\gamma$, $T_0$, $q$, $\xi$, and $v_{sys}$. 

We also allow the viewing geometry of the system, the inclination and position angle to vary in our fit. We use the 3D radiative transfer modelling package \texttt{RADMC-3D} \citep{Dullemond2012} to calculate the molecular level populations in each cell and produce synthetic ${}^{12}$CO (2--1) channel maps for a given set of model parameters. Those synthetic channel maps are Fourier Transformed and fit directly to the visibilities using the MCMC fitting package \texttt{emcee} \citep{FM13}, with uniform priors for all parameters.

\subsection{Hydrodynamic and Near-Infrared Radiative Transfer Modelling}

To investigate the structures induced by the companion on the disk around the primary star, we utilize combined hydrodynamic and radiative transfer simulations, following the strategy in \cite{Dong2016b}. We utilize the three-dimensional hydrodynamic package, \texttt{PEnGUIn} \citep{Fung2015} and the radiative transfer package, \texttt{HOCHUNK3D} \citep{Whitney2013} to compute the time-dependent density evolution of the disk and its resulting appearance in near-infrared scattered light. We compute synthetic H-band (1.65 $\mu$m) images assuming that the disk is composed of primarily interstellar medium grains \citep{Kim1994}. These grains contain silicate, graphite, and amorphous carbon, and their size distribution is represented by a smooth power law distribution in the range of 0.02$-$0.25 $\mu$m followed by an exponential cut off beyond 0.25 $\mu$m. The anisotropic scattering phase function is approximated using the Henyey-Greenstein function \citep{Henyey1941}. The optical properties can be found in Dong et al. (2012, Figure 2).\footnote{While a better match to the observed scattered light profile of the disk may be achieved by fine-tuning the dust constituents and anisotropic scattering properties, such work is beyond the scope of this study that is aimed only at a first-order qualitative match to the observed disk structure.}

In the hydrodynamic model, the companion is orbiting in the same direction and plane as the disk on a circular orbit with a semi-major axis of 100 AU, which is consistent with our derived orbital parameters (\S4.2). The three-dimensional disk density profile was computed for one-hundred companion orbits, and the final density distribution was input into the radiative transfer software to simulate how the disk would appear in $H$-band scattered light (1.65 $\mu$m). We reproduce the model three times with an inclination of $25^\circ, 30^\circ$, and $35^\circ$ from face-on, to show the resulting disk structures as viewed across the range of plausible disk inclinations. We note that the companion-to-star mass ratio has been lowered from 1:6 as in our previous study \citep{Dong2016a} to 1:10, as we now adopt $M_{\rm HD~100435B}=0.2M_\odot$ (Collins et al. 2009) instead of $0.3M_\odot$ (Chen et al. 2006).
\section{Results}

\subsection{Astrometric Catalog from 2003-2017}

We provide an astrometric catalog of six epochs spread over fourteen years of high-contrast adaptive optics observations with VLT/NACO, VLT/SPHERE, and MagAO/Clio-2 (Table 2). These measurements constitute all of the high-fidelity astrometry on HD 100453B that is known to us. Notably, other observations exist from these and other facilities throughout this timespan, but are excluded from our catalog due to their lack of extened baseline and poorer astrometric quality. Here we briefly summarize the other available data known to us. \cite{Collins2009} provide an additional measurement with HST/ACS that is consistent with the NACO measurements, although the astrometric uncertainties are 4-5$\times$ larger. Other publicly available data from NACO exists as well, however, these observations were carried out with a coronagraph\footnote{Without the benefit of satellite spots, similar to coronagraphic SPHERE calibrations.}, and thus suffer from uncertainty in determining the location of the primary star behind the coronagraph. Thus, we have chosen to exclude these measurements from our catalog and proceeding analysis.

\begin{deluxetable}{ccccc}
\tabletypesize{\scriptsize}
\tablecaption{HD 100453AB Astrometric Catalog}
\tablewidth{0pt}
\tablehead{
\colhead{Date} & \colhead{Facility} & \colhead{Separation} & \colhead{PA} & \colhead{Reference}
}
\startdata
2003-06-02 & VLT/NACO & 1$\farcs$049$\pm$0$\farcs$0070 & 127.23$^\circ\pm$0.33$^\circ$ & a \\
2006-06-22 & VLT/NACO & 1$\farcs$042$\pm$0$\farcs$0050 & 128.26$^\circ\pm$0.30$^\circ$ & a \\
2015-04-10 & VLT/SPHERE & 1$\farcs$047$\pm$0$\farcs$0031 & 131.63$^\circ\pm$0.19$^\circ$ & This work \\
2016-01-21 & VLT/SPHERE & 1$\farcs$056$\pm$0$\farcs$0030 & 131.95$^\circ\pm$0.19$^\circ$ & This work \\
2016-01-23 & VLT/SPHERE & 1$\farcs$053$\pm$0$\farcs$0030 & 131.95$^\circ\pm$0.19$^\circ$ & This work \\
2017-02-17 & MagAO/Clio-2 & 1$\farcs$056$\pm$0$\farcs$0054 & 132.32$^\circ\pm$0.44$^\circ$ & This work \\

\enddata
\tablenotetext{a}{Data from \cite{Collins2009} adjusted according to the astrometric calibrations in \cite{Chauvin2010}. }
\end{deluxetable}


\subsection{Orbital Parameters of HD 100453AB}

The main focus of this work is to establish the orbital parameters of the HD 100453AB binary, and to relate these to the probable mechanism behind the spiral arms in the disk. The results of our orbit fitting (described in \S 3.1) are shown in Figures 4 \& 5. The histograms of remaining orbital parameters (node, longitude of periastron, and time of periastron passage) are shown in in Appendix A. We note that our six-parameter model\footnote{With the seventh parameter, the orbital period, linked to semi-major axis by mass.} is over-fitting our six available data points $-$ i.e. the best-fit model is not necessarily representative of the true orbit, but likely represents the particular noise realization in the data. While this remains a limitation of our model, the problem is mitigated by our bootstrapping method of parameter retrievals, which randomizes the noise in the data while assembling a large volume of orbital parameter retrievals. This method results in useful constraints on the semi-major axis, eccentricity, and inclination, while leaving the node and longitude of periastron relatively unconstrained (see Appendix A). In other words, with the data at hand we are able to place sufficient constraints on the orbital geometry, while leaving the exact orbit within this parameter space unconstrained.

\begin{figure}[htpb]
\figurenum{4}
\epsscale{1.}
\plotone{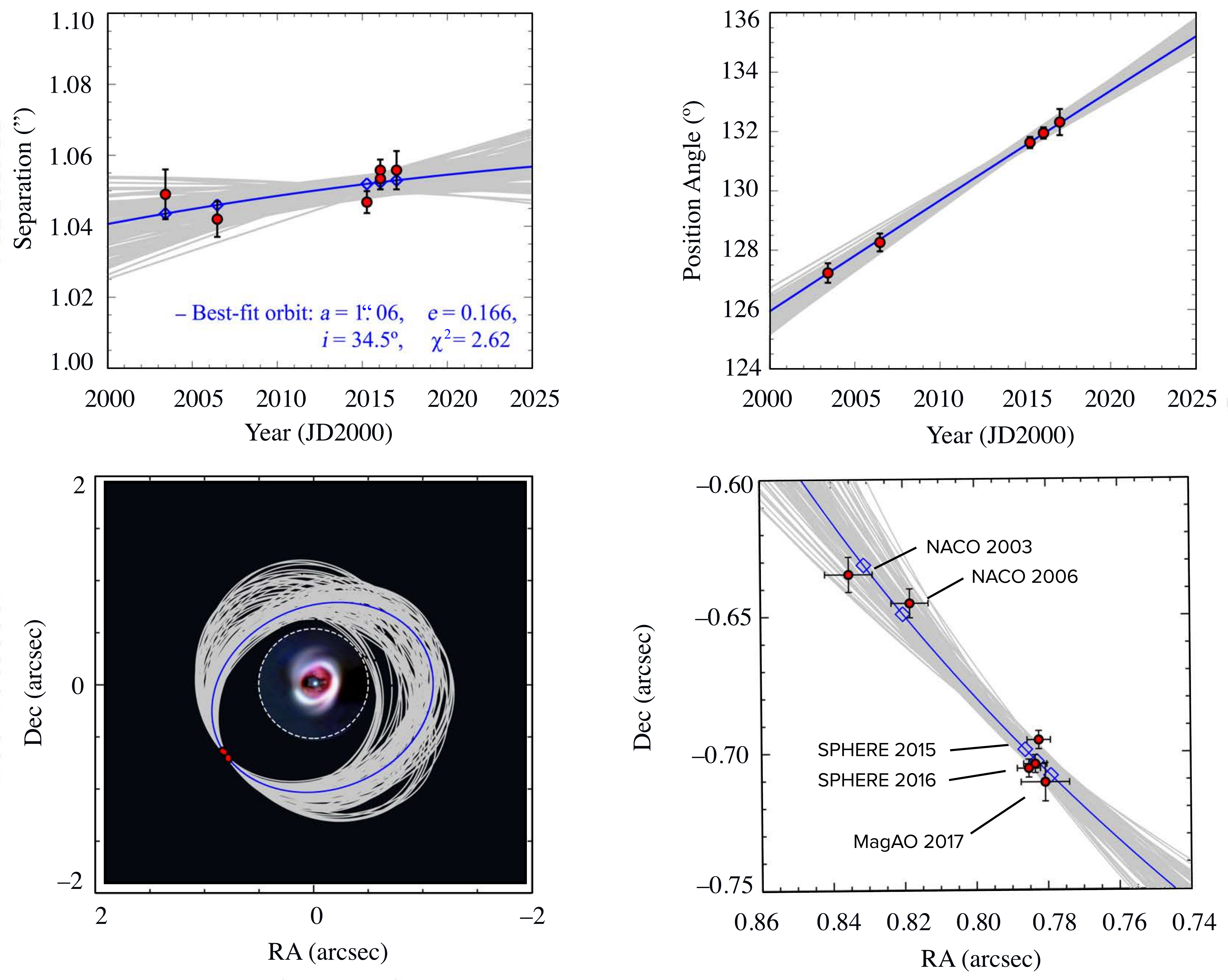}
\caption{Results from our orbit fit for HD 100453AB. The blue curves show the best fit orbit while those plotted in gray represent one hundred orbits drawn at random from 25,265 trials to show the range of plausible orbits. In the bottom left panel, the scattered light image of the disk from \cite{Wagner2015} is shown as an inset (dashed region) to illustrate the proximity of the companion's orbit to the disk. Since the true location of the system's barycenter is unknown, the motions of the system are shown with the A-star as a common central reference point.}
\end{figure}

We establish 1$\sigma$ confidence intervals\footnote{Given by the mean and standard deviation of our complete set of retrieved orbital parameters.} on the orbital parameters \textit{a}=1$\farcs$06$\pm0\farcs09$, \textit{e}$=0.17\pm0.07$, and \textit{i}=$32.5^{\circ}\pm6.5^\circ$. Notably, the mechanisms proposed to generate spiral arms without a strong influence from the companion depend on parameters of the disk physics that are not well constrained (e.g., cooling timescales in the case of shadow-generated spiral arms, as in \citealt{Benisty2017}). However, the effects due to gravity, in the case of a companion on a relatively close orbit, cannot be ignored. Conversely, if the orbit is oriented such that it is masking a larger pericenter, then the gravitational effects of the binary may be negligible. Consistent with the former scenario, we find that the companion is on an orbit that is at most mildly eccentric and with a semi-major axis that is similar to the measured projected separation (a$\sim$105 au). The inclination of the companion's orbit is co-planar with the disk to within measurable limits. The effects of this orbital configuration on the disk structures are discussed in \S 5.1.

\begin{figure}[htpb]
\figurenum{5}
\epsscale{1.}
\plotone{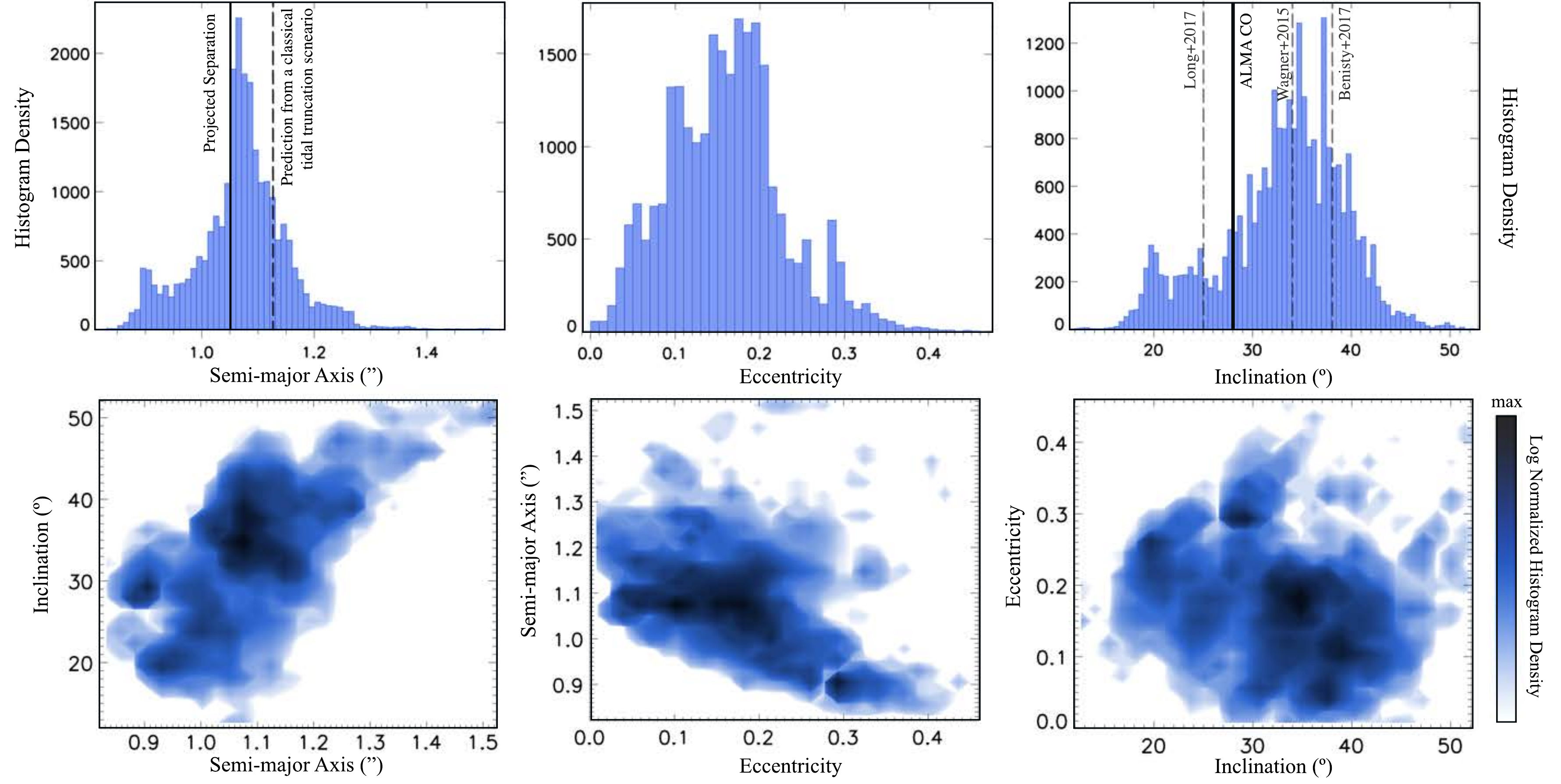}
\caption{\footnotesize One$-$ and two-dimensional histogram densities from our orbital parameter retrievals. In the top left panel (semi-major axis) the solid black line corresponds to the projected separation of the companion, while the dashed line corresponds to the predicted orbital semi-major axis of the companion, assuming a strictly co-planar and circular orbit from classical tidal truncation theory (e.g., \citealt{Holman1999}). In the top right panel (inclination), the dashed lines represent estimates of the disk inclination from near-infrared scattered light images, and the solid line represents the inclination derived by our ALMA $^{12}$CO $J$=2-1 kinematic modelling.}
\end{figure}

\subsection{Disk Inclination}
\subsubsection{NIR Images and Radiative Transfer SED Modelling}

To understand the effects of HD 100453B on the disk, it is first necessary to establish the mutual inclinations in the system. The inclination of the disk has been previously explored through fitting an oval to the ring of emission and assuming a circular geometry (zero eccentricity) to estimate the degree of inclination. Using this strategy, \cite{Wagner2015} found a disk inclination of $\sim$34$^\circ$ from face-on. In agreement with this value, \cite{Benisty2017} utilized a similar method on their polarimetric data and found a disk inclination of $\sim$38$^\circ$ from face-on. 

However, these estimates do not take into account the fact that the disk has a non-negligible scale height and a wall-like structure at the radius of the ring, and that scattering effeciency depends on the scattering angle which is different for the opposing sides, and hence the brightest region of the far-side of the disk will in fact be the directly illuminated edge of the gap, not the upper region of the disk as seen on the near-side. \cite{Long2017} take this into account via modelling the disk to fit both the SED and the images, repeating the inclination estimate on radiative transfer images from their models convolved to the same angular resolution of VLT/SPHERE. They find a somewhat smaller (though broadly consistent) inclination of 25$\pm$10$^\circ$ from face-on. 

\subsubsection{ALMA Disk Kinematic Modelling}

Following the modelling described in \S3.2, we derive the general disk properties for our assumed model geometry $-$ i.e. a circular, flaring disk with a radial power-law distribution for the disk surface density and an exponential vertical cutoff. Given that the disk is poorly resolved, our model mostly reflects the properties of the outer disk, since this region contains the bulk of the disk surface area and CO emission. The inclination obtained from ALMA for the outer disk around HD 100453A ($\sim$28$^{\circ}$) is consistent with the independent measurements from the NIR images (Figure 5). The best-fit model parameters and uncertainties are shown in Table 3.  The channel 0 and channel 1 moment maps for the data and model are shown in Figure 6, while the model channel maps and residuals from the model-data fit are shown in Appendix B.

\begin{deluxetable}{ccc}
\tabletypesize{\scriptsize}
\tablecaption{Model Parameters}
\tablewidth{0pt}
\tablehead{
\colhead{Parameter} & \colhead{Value} & \colhead{Uncertainty} 
}
\startdata
$M_{Disk}$ (M$_{\odot}$)& 7.0$\times 10^{-6}$ &  1.0$\times 10^{-6}$  \\
$r_{c}$ (AU) & 27 & 1\\
$T_{0}$ (K) & 511 & 72\\
$a_{turb}$ (km/s) & 0.92 & 0.03\\
$v_{sys}$ (km/s) & 5.25 & 0.01\\
$i_{Disk}$ ($^{\circ}$) & 27.6 & 0.2\\
$PA$ ($^{\circ}$) & 55.0 & 0.5\\
$q$ & 0.37 & 0.07\\


\enddata
\tablecomments{The quoted uncertainties do not take into account the uncertainty inherent in the model assumptions.}
\end{deluxetable}

Given the complexity of the inner structures in HD 100453, our simple model cannot capture all of the disk features that result in the observed CO emission profile. The disk ring has two dips at relatively symmetric positions that are most likely shadows produced by a highly misaligned inner disk, as in the case of HD 142527 \citep{Marino2015}. The presence of such an inner disk can alter the rotation pattern in the gas by introducing a distinctive “twist” in the moment 1 map (\citealt{Casassus2015}, \citealt{Facchini2017}). While the ALMA dataset used in this work does not have the angular resolution to clearly reveal such a pattern, it may well be present in the data, and cause the moment 1 map to deviate from the typical “butterfly” pattern from a co-planer, inclined disk. Detailed modeling of the ALMA data taking this effect into account, which is beyond the scope of this paper, is needed to more accurately pin down the disk parameters. Also, while a better match to the outer extent of the disk may be achieved through more detailed modeling, namely through a more selective prior distribution for $r_{c}$, the presence of the spiral arms complicates such a selection. Since our study is not primarily concerned with this property of the disk, we have not made an attempt to generate such a model. A more detailed model of the disk will be presented in van der Plas et al, in prep.

\begin{figure}[htpb]
\figurenum{6}
\epsscale{1.}
\plotone{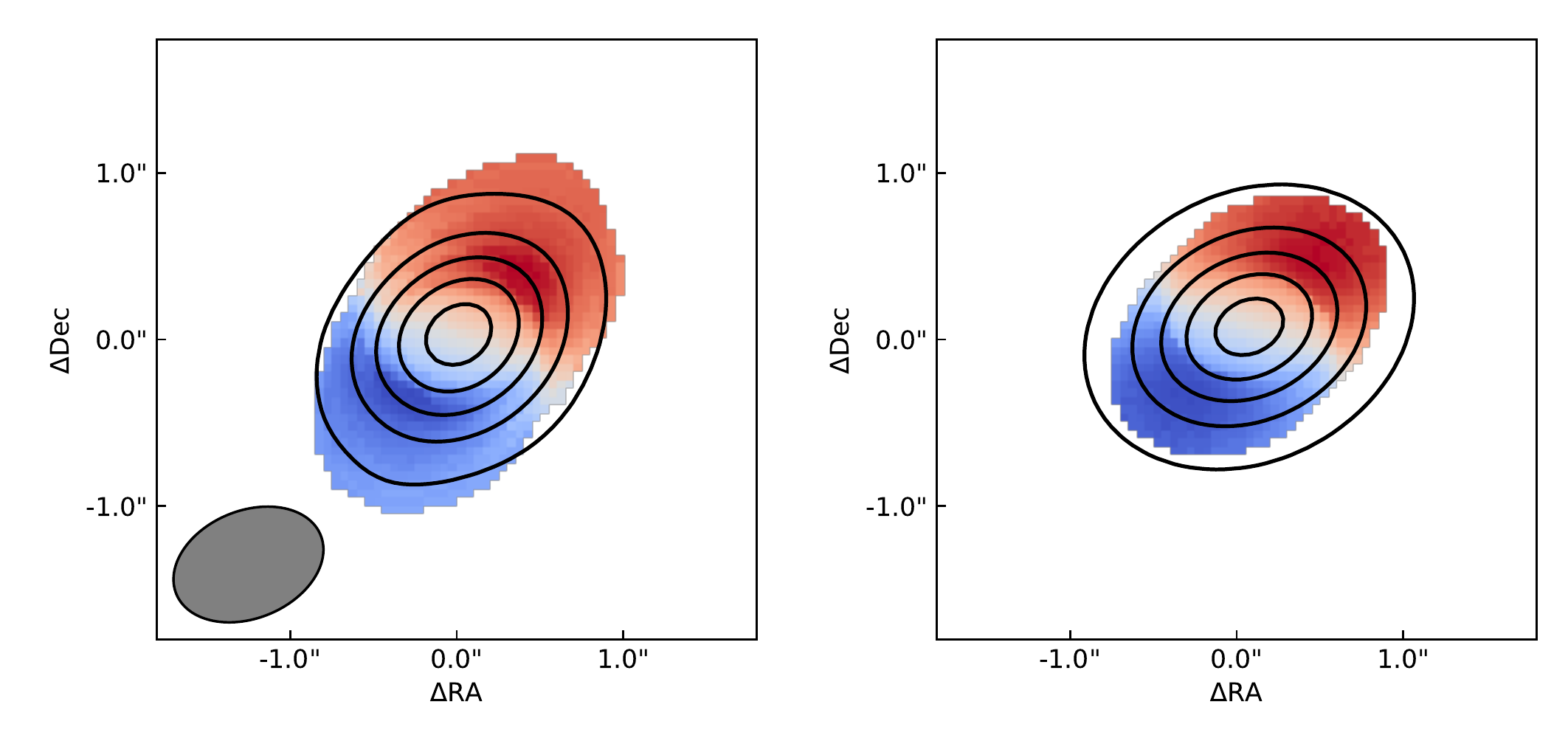}
\caption{ALMA CO moment zeroth and first moment maps for the data taken on 2016 April 23 (left) and our model disk (right). The beam size is shown as the inset grey region, and the synthetic beam size of the model image is identical. The cause of the mismatch between the size of the disk and the model is likely two-fold: due to lower noise in the model image, and due to its lack of spiral arms. This is acceptable for our purposes, as we are primarily interested in the inclination of the disk and not its exact spatial extent.  }
\end{figure}

\subsection{Hydrodynamic + Radiative Transfer Simulations}

\begin{figure}[htpb]
\figurenum{7}
\epsscale{1.}
\plotone{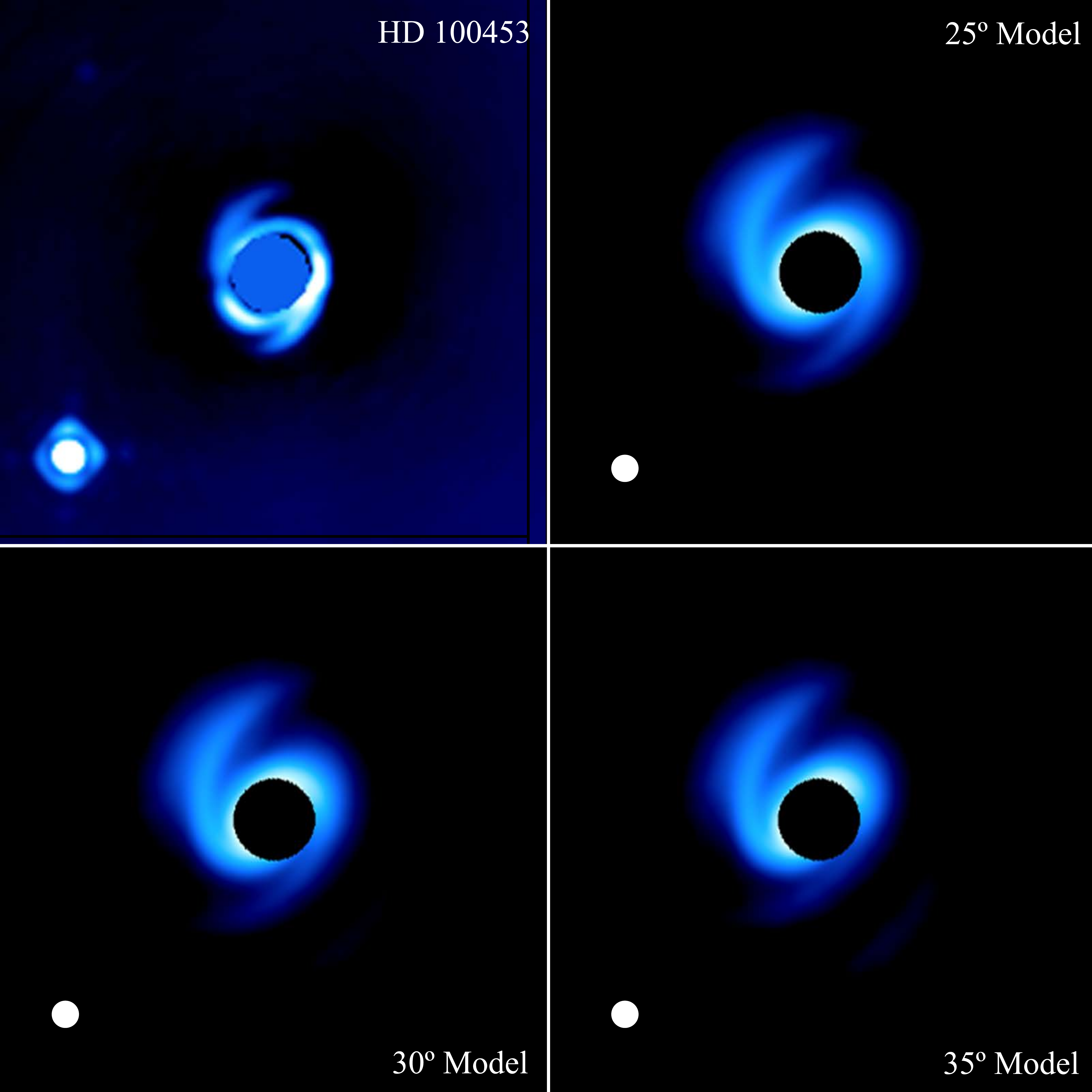}
\caption{VLT/SPHERE NIR scattered light image of HD 100453 (top left) and hydrodynamic + radiative transfer simulations covering the range of possible inclinations. The simulations exhibit a prominent two armed spiral structure induced by the companion, suggesting a similar origin for the actual structures in HD 100453. The intensity of the images is shown on a logarithmic scale (arbitrary units) for clarity in the faint features. The gap interior to 20 AU and the warp in the inner disk are not included in the models.}
\end{figure}

Motivated by the new constraints provided by the orbital fit and ALMA gas kinematics, we utilized combined hydrodynamic and radiative transfer simulations (\S3.3) to compute the three-dimensional disk density evolution of an initially axial-symmetric disk perturbed by an outer companion for one hundred orbits. We then generated synthetic NIR scattered light images to be compared to the observations. The images from our three simulations, in which we varied the system inclination throughout a plausible range of 25-35$^{\circ}$, are shown in Figure 7. The model images are rotated such that the companion is at its known PA=135$^\circ$. Throughout the plausible range of inclinations we find that the companion induces a prominent two-armed spiral structure that is qualitatively similar to the observed spirals in HD 100453. We find the best match to the data arising from the models with a system inclination of 25-30$^{\circ}$, consistent with estimates on the disk inclination from our ALMA CO modelling and from our orbital fit, both of which suggest a common inclination for the outer disk and companion of $\sim$30$^{o}$. The separation of the spiral arms from one another, their pitch angle, and originating locations among the disk ring are reproduced by the model. Similar to the observations, the primary arm is pointed toward the companion, with the secondary arm opposite. The bottom side of the disk is also visible to the Southwest (see Appendix C), which is consistent with the polarized intensity imaging of \cite{Benisty2017}. The implications and limitations of these results are discussed further in \S5.1.

\section{Discussion}

\subsection{Effects on the Disk Structure from HD 100453B}

Our primary goal in this work is to investigate the orbit of the HD 100453AB binary and its resulting effects on the protoplanetary disk structure. With our fourteen year baseline of astrometry, we are able to provide the first fit to the orbit of these stars. We combine this with estimates on the disk geometry from the literature and with our analysis of public ALMA data to constrain the inclination and major axis of the disk. We find that the binary's orbit is roughly co-planar with the disk, and at most mildly eccentric (e$\lesssim$0.3). Furthermore, the companion's orbital semi-major axis is 3-4$\times$ greater than the observed outer disk radius ($\sim$40 au). This is, as expected, consistent with the classical disk truncation scenario (e.g., \citealt{Artymowicz1994}, \citealt{Holman1999}). The former study calculated the size of the circumprimary disk as it is dynamically truncated by a 1:10 mass-ratio companion on a variety of orbits. Their results suggested that for disk scale height and viscosity plausible for protoplanetary disks, such a companion on an e=0.2 orbit could truncate the circumprimary disk to 30-40\% of the binary semi-major axis. This is in excellent agreement with the detected disk size of HD 100453, and combined with the atypical small size of the disk, is strong evidence that it is indeed dynamically sculpted by the companion. 

We find that the pair of spiral arms generated by the companion in our simulations are qualitatively well-matched to the spiral arms in the observed disk structure, and thus we suggest that any additional effects (e.g. differential thermal and radiation pressure in the shadowed regions) are not necessary to explain their origin. The implications of this result for the other two ``grand-design" (two-armed) spiral protoplanetary disks are discussed in \S5.2. We note that comparing with the observation, the disk size in our simulation (Figure 7) appears to be $\sim$30\% larger. This is probably caused by two factors. First, due to the high computational cost, we only evolve the simulation for 100 companion orbits. In contrast, HD 100453AB is 10$\pm$2 Myr old (\citealt{Collins2009}), corresponding to 12,000 orbits. While 100 orbits is sufficient to truncate the bulk part of the disk, it will be further truncated by a small amount between 100 and 10,000 orbits, thus achieving a better agreement with the observation. Second, \cite{Artymowicz1994} showed that the truncation radius modestly depends on the scale height (h/r) and viscosity ($\alpha$) in the disk. In our simulation we parametrized the disk to have $h/r$ = 0.15 and $\alpha$ = 10$^{-3}$, which are not expected to be in exact agreement with the scale height and viscosity in HD 100453. Therefore, we do not expect an exact reproduction of the observed disk morphology with our simulation.

Finally, we note that our modelling approach has several other limitations, which we briefly discuss here. First, we do not include the possible small inclination of the companion relative to the disk ($\leq 10^{\circ}$ with $1\sigma$ confidence; or $\leq 20^{\circ}$ with $2\sigma$ confidence), as well as its small eccentricity (0.15$\pm$0.7 within 1$\sigma$, or $\lesssim 0.3$ within 2$\sigma$). However, we do not expect these to result in any major differences in disk structure of the models. Indeed, this choice is justified since due to tidal dissipation within the viscous disk we expect the binary to be on a relatively circular orbit and to be co-aligned with the disk (e.g., \citealt{Pap1995}, \citealt{Bate2000}, \citealt{Lubow2000}). These dynamical effects are discussed further in \S5.3.  Finally, the scattering phase functions of the grains in the model and grains in the real disk bear noticeable differences, notably in the forward scattering of the near (Southwest) side of the disk. While more realistic grain populations and scattering properties could improve the quantitative match of the disk flux and contrast of the spiral arms, such work is beyond the scope of this current study.

\subsection{Implications for Other Disks}

While it appears clear that the ``grand-design" spiral arms in HD 100453 are driven by the low-mass stellar companion, the nature of the other such spiral disks (SAO~206462: \citealt{Muto2012}, MWC~758: \citealt{Grady2013}, LkH$\alpha$ 330: \citealt{Akiyama2016}) remains unclear. Binary companions as massive as HD 100453B would have been detected in these systems, if they were present. Therefore, if these spiral structures are generated in the same way as those in HD 100453, then the companions must be of sufficiently low mass to avoid detection in direct imaging ($\lesssim$ 2$-$4 $M_{J}$ at $\geq$0$\farcs$6 for SAO 206462 assuming hot-start models, \citealt{Maire2017}). If such low-mass companions are responsible for the observed spiral structures, then they must be on the lower end of masses predicted to be capable of this feat \citep{Fung2015b}. Alternatively, it is possible that the hot-start evolutionary tracks over-predict the luminosity of young giant planets, which would explain the non-detection of planets at the hot-start mass-luminosity predictions for the bodies driving the spiral arms in SAO 206462 and MWC 758.

\subsection{The Distribution of Angular Moment in HD 100453: Evidence for a Primordial Alignment}

While the orbit of the binary appears to be closely co-planar with the outer disk, in which it drives spiral density waves, an additional disk component with apparently distinct morphology exists closer in to the primary star. This inner disk is highly misaligned ($\sim$45-70$^{\circ}$) with respect to the plane of the outer disk (and binary), as evidenced by the prominent shadows cast by this feature on the outer disk (\citealt{Benisty2017}, \citealt{Long2017}), similar to the shadows cast by the misaligned inner disk in HD 142527 \citep{Marino2015}. The origin of this misalignment is presently unknown, though a possible scenario involves a companion orbiting within the gap and generating the misalignment, similar to the observed configuration of HD 142527 \citep{Close2014}. However, since HD 100453B is on an orbit that is altering the structure of the outer disk, it is reasonable to speculate that it may have also altered the inclination of the outer disk to more closely resemble that of its present orbit. In this case, the orientation of the inner disk and the rotational plane of the primary star, which may be relatively unaffected by the binary, may preserve the direction of the original angular momentum of the parent cloud which formed the HD 100453 system. 

Indeed, presuming that the disk and binary shared some initial significantly non-zero mutual inclination, simulations predict that given the age of the system, we should expect the outer disk and binary to be co-aligned due to tidal and viscous dissipation within the disk (e.g., \citealt{Pap1995}, \citealt{Bate2000}, \citealt{Lubow2000}). If the mutual inclination is larger than the disk opening angle, then due to hydrodynamic instabilities the alignment timescale is comparable to the disk precession timescale, which for typical protostellar disk parameters is on the order of $\sim$20 binary orbits \citep{Bate2000}, or $\sim$10$^{4}$ years. This is much smaller than the system's age of 10 Myr, and thus the mutual inclination should be less than the disk opening angle. In this case, the timescale for further alignment is on the order of the disk viscous timescale. Assuming conservative estimates for the viscosity parameter, $\alpha$=0.001, and disk aspect ratio, $h/r$=0.05, the viscous timescale is $\sim$10 Myr, which is comparable to the age of the system. Thus, the disk around HD 100453A should have had sufficient time to align itself to the orbital plane of the binary, and thus possibly does not represent its initial direction of angular moment (e.g., if HD 100453B did not form in the system, and was subsequently captured during close passage). 

To test whether there may have been an initial misalignment of the disk around HD 100453A and the orbit of HD 100453B, we consider the obliquity of HD 100453A from its measured rate of rotation. If its angular momentum is aligned with the binary (and thus the outer disk), then we postulate that the system likely formed from the same parent cloud and inherited a common direction of angular momentum. On the other hand, if the primary star's rotational plane is significantly misaligned with respect to these outer components, as is the inner disk, then perhaps these inner components represent the initial angular momentum of the cloud that formed HD 100453A and its disk, leaving the misalignment of the companion to require further explanation. \cite{Guimaraes2006} measured the rotation of HD 100453A, $v~sini = 48\pm 2$ km/s. For comparison, \cite{Zorec2012} found that the distribution of equatorial velocities for stars between 1.6-2.0 M$_{\odot}$ peaks at $\sim$150 km/s with a 1$\sigma$ spread of $\sim$50 km/s. Assuming an equatorial velocity for HD 100453A of 100, 150, and 200 km/s, the measured $v~sini $ corresponds to an inclination of 15$^{\circ}$, 20$^{\circ}$, and 30$^{\circ}$ from face-on, respectively, which more closely resembles the inclination of the outer disk and companion. Thus, we suggest that the inner disk, and not the outer disk, is the misaligned body in the system, and that the primary star's rotation, the orbit of the outer disk, and the orbit of the companion likely reflect the initial angular momentum of the parent body that formed the system.


\section{Summary and Conclusions}

We have obtained new VLT/SPHERE and Magellan/MagAO observations of HD 100453 in $H23$, $K12$, and $L^{\prime}$ filters (\S 2.1 \& \S 2.2). These data establish a fourteen year baseline with previous VLT/NACO adaptive optics imaging, and enable the first Keplerian orbital fit to the stellar motions. We fit the six epochs of astrometric data to a grid of models and find a best fit with orbital parameters of \textit{a}=1$\farcs$08 (111 au @ 103 pc),\textit{ e}=0.12, and \textit{i}=32.6$^{\circ}$. From 25,265 parameter retrievals with our bootstrapping method, we establish 1$\sigma$ confidence intervals on the orbital parameters \textit{a}=1$\farcs$06$\pm0\farcs09$, \textit{e}$=0.17\pm0.07$, and \textit{i}=$32.5^{\circ}\pm6.5^\circ$. We utilize publicly available ALMA $^{12}$CO observations to model the outer disk properties, and find a best-fit with an outer disk inclination of $\sim$28$^{\circ}$ from face-on. With the combined knowledge that the companion is on a near-circular orbit that is 3-4$\times$ larger than the disk, to which it is also co-planar, we find that HD 100453 represents a classical scenario of a circumstellar disk that is truncated by an external companion (e.g., \citealt{Artymowicz1994}, \citealt{Holman1999}). The full range of retrieved orbital parameters (Figure 5 and \S 4.2), and disk geometry (\S 2.5 \& \S 4.3.2), are fully consistent with this scenario, which is further supported by the atypical small size of the disk around HD 100453A. 

We utilized hydrodynamic and radiative transfer simulations to model the disk structures induced by the system's binary nature (\S 4.4). We find that the companion generates a two-armed spiral structure in the simulated disk that is well-matched to the observed general properties of the spiral arms in HD 100453, consistent with preliminary findings of \cite{Dong2016a} . Thus, any dynamical model of the disk structure must include the effects of the companion's gravitational perturbations in order to accurately match the dynamics of the system. We suggest that the other proposed spiral-arm-producing mechanisms (e.g., \citealt{Montesinos2016}, \citealt{Kama2016}) are not responsible for the ``grand-design" spiral structure of HD 100453. Finally, we note that the inclinations of the various system components, with the exception of the inner disk, are consistent with a co-planar system. Future work is required to elucidate the system configuration close to the primary star, including the presently enigmatic origin of the gap and the misaligned inner disk, which will further help to reveal the system's rich and complex dynamical history.

\section{Acknowledgments}

The authors wish to express their gratitude toward Kaitlin Kratter and Myriam Benisty for useful conversations and feedback on this manuscript, as well as toward the anonymous referee who provided numerous helpful suggestions and improvements, and to Karen Collins, Roy van Boekel, and Xuepeng Chen for providing additional information on the astrometric calibrations of the previously published NACO data. The results reported herein benefited from collaborations and/or information exchange within NASA's Nexus for Exoplanet System Science (NExSS) research coordination network sponsored by NASA's Science Mission Directorate. This paper makes use of the following ALMA data: ADS/JAO.ALMA\#2015.1.00192.S. ALMA is a partnership of ESO (representing its member states), NSF (USA) and NINS (Japan), together with NRC (Canada), NSC and ASIAA (Taiwan), and KASI (Republic of Korea), in cooperation with the Republic of Chile. The Joint ALMA Observatory is operated by ESO, AUI/NRAO, and NAOJ. The National Radio Astronomy Observatory is a facility of the National Science Foundation operated under cooperative agreement by Associated Universities, Inc. This paper includes data gathered with the 6.5 meter Magellan Telescopes located at Las Campanas Observatory, Chile. Based on observations made with ESO Telescopes at the La Silla Paranal Observatory under programme ID 095.C-0389(A). KRW is supported by the National Science Foundation Graduate Research Fellowship Program under Grant No. 2015209499. KMM's and LMC's work is supported by the NASA Exoplanets Research Program (XRP) by cooperative agreement NNX16AD44G.


\section{Appendix}
\appendix

\section{Orbital Parameter Histograms (Continued)}

\begin{figure}[htpb]
\figurenum{A1}
\epsscale{1.}
\plotone{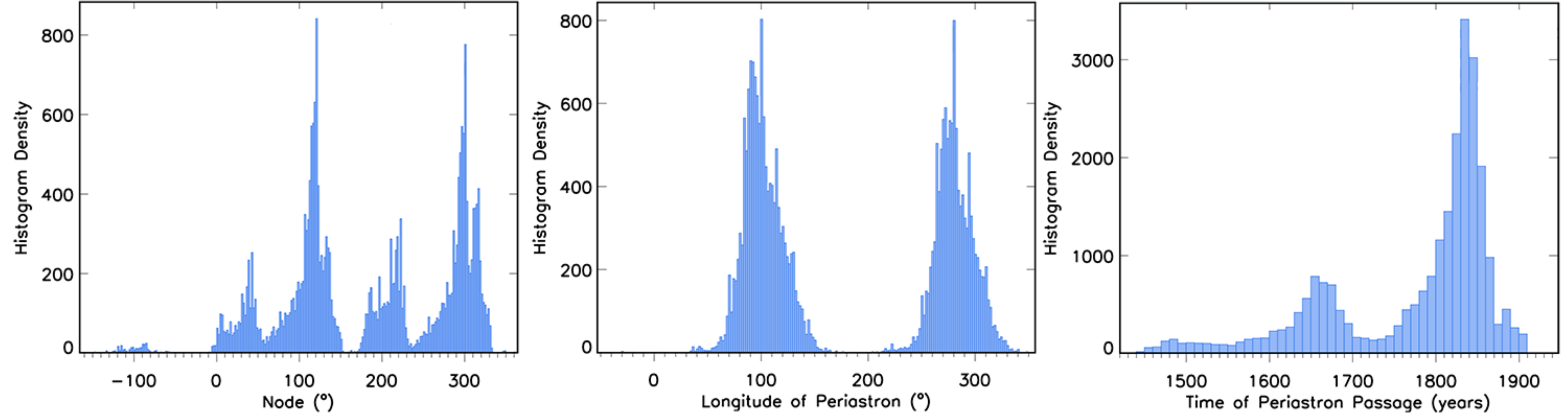}
\caption{One-dimensional histograms of retrieved node, longitude of periastron, and time of periastron passage for HD 100453B. Note that since the curvature of the orbit has not yet been well determined, the orientation of the node bears a four-peaked structure with $\sim$90$^{\circ}$ spacing. Similarly, the longitude of periastron displays a double peaked structure, with peaks spaced by $\sim$180$^{\circ}$. In other words, the unconstrained nature of these parameters shows that we have not determined the exact orbit of the companion, but rather the distribution of $a, e, i$ in \S4.2 constrain the type of orbit, and thus its proximity and dynamical influence on the disk.}
\end{figure}

\section{Model CO Channel Maps and Residuals}

\begin{figure}[htpb]
\figurenum{B1}
\epsscale{1.}
\plotone{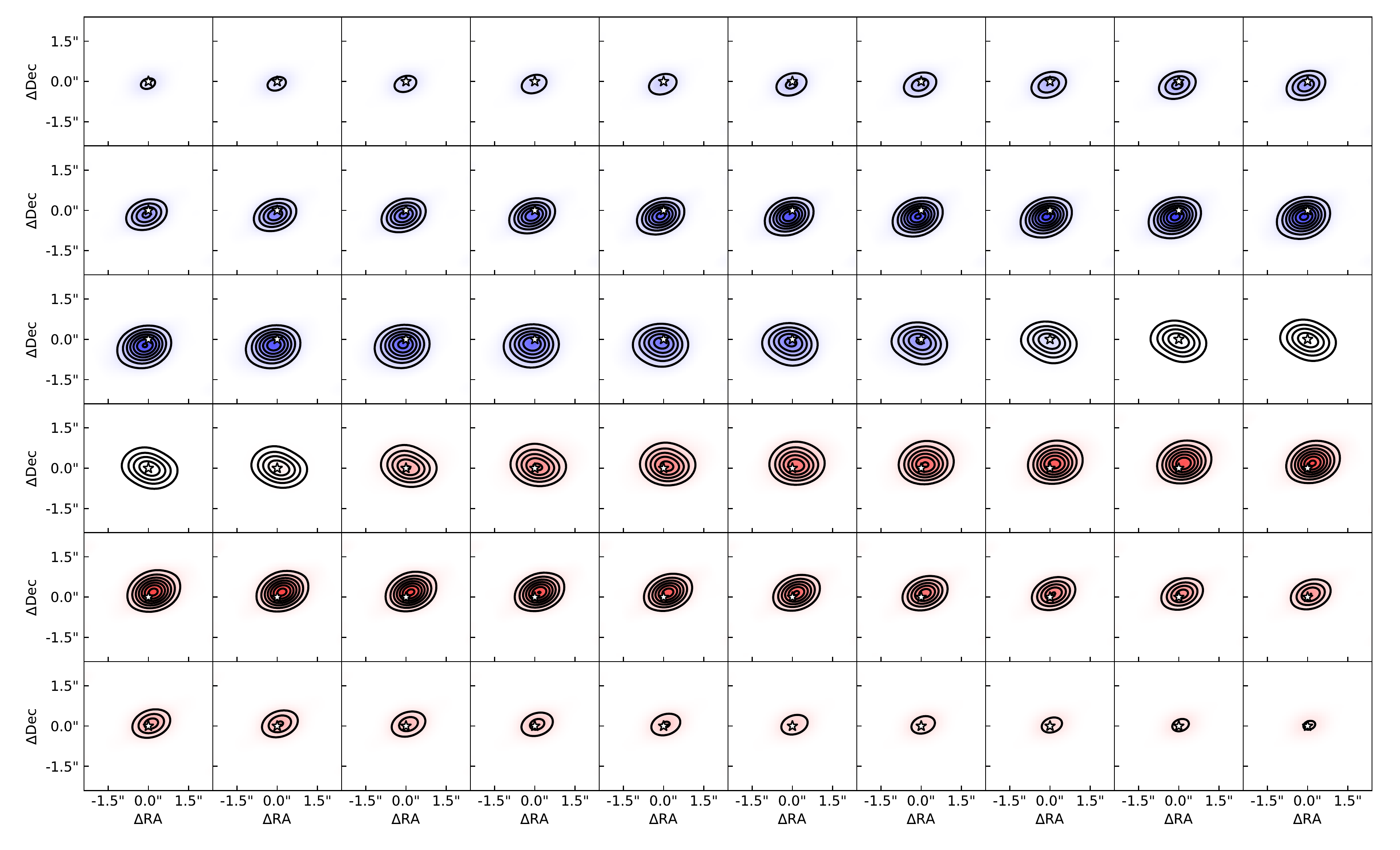}
\caption{Model channel maps of $^{12}$CO 2--1 emission in the disk surrounding HD 100453A. The contours represent integer multiples of the $3\sigma$ detection limit ($\sigma$=8.1 mJy/beam).}
\end{figure}

\begin{figure}[htpb]
\figurenum{B2}
\epsscale{1.}
\plotone{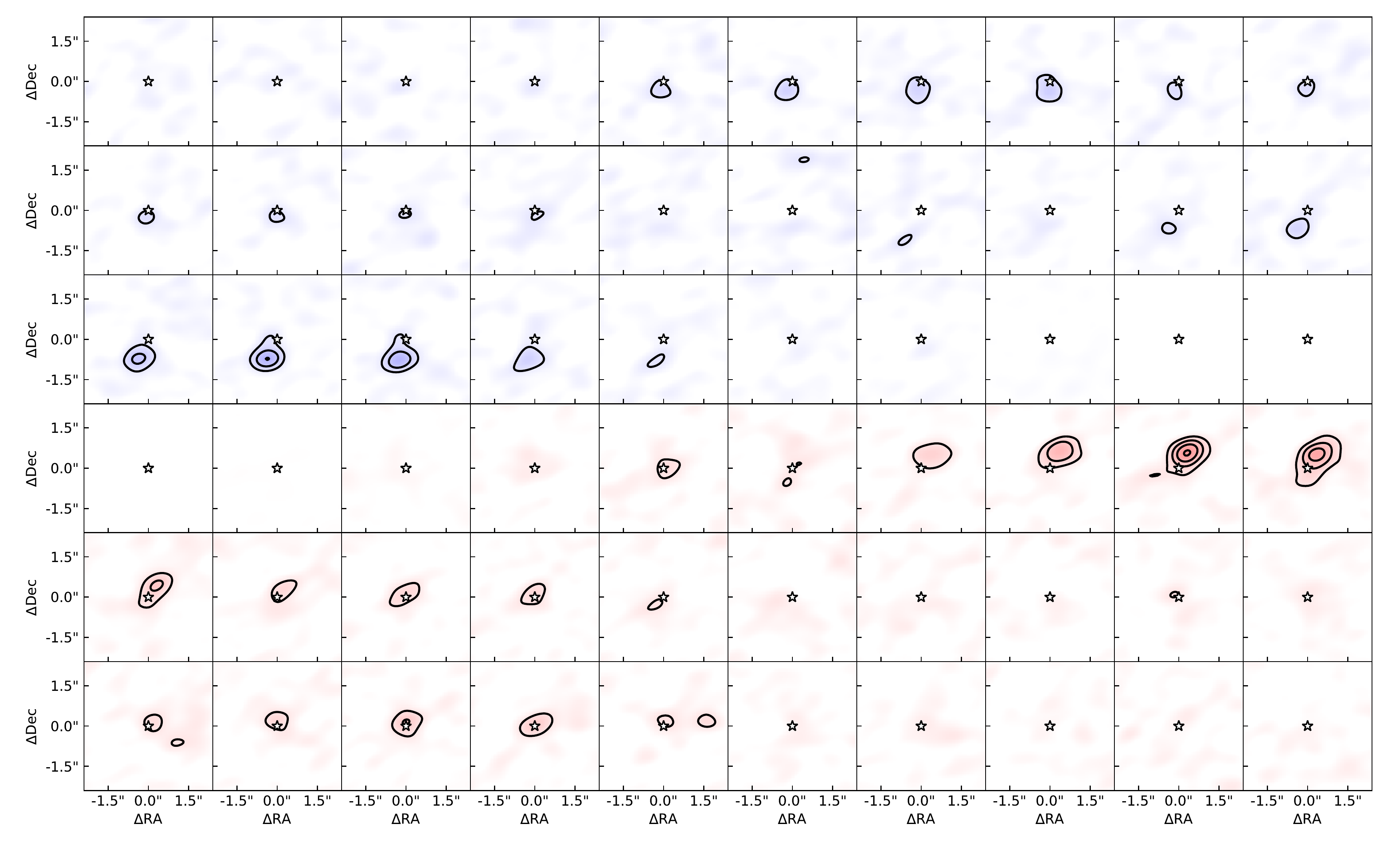}
\caption{Residual channel maps of $^{12}$CO 2--1 emission in the disk surrounding HD 100453A generating by subtracting our model from ALMA observations taken on 2016 April 23. The contours represent integer multiples of the $3\sigma$ detection limit ($\sigma$=8.1 mJy/beam).}
\end{figure}

\section{Disk Bottom Side Data and Model Comparison}

\begin{figure}[htpb]
\figurenum{C1}
\epsscale{1.}
\plotone{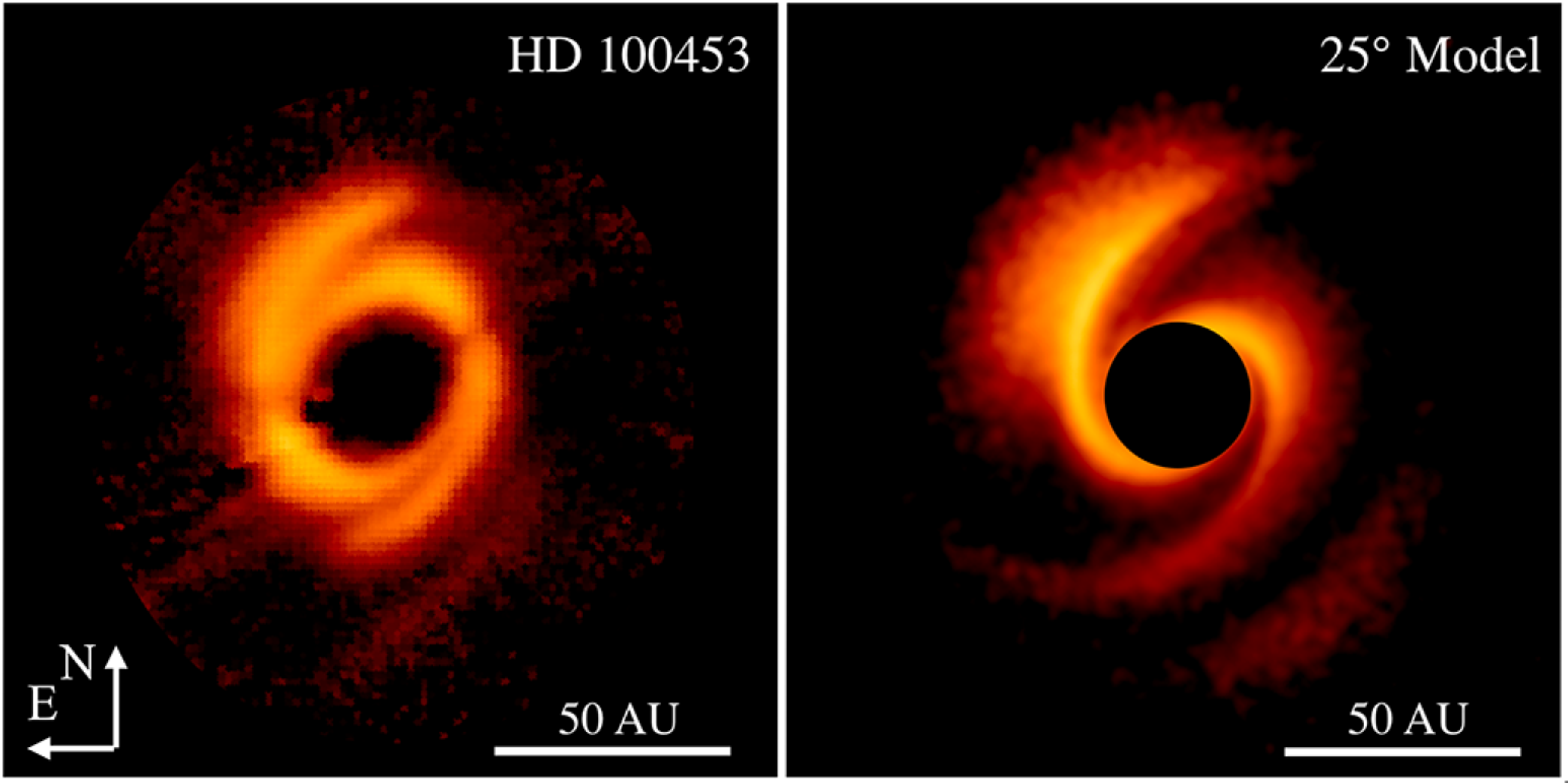}
\caption{\textit{Left}: VLT/SPHERE $J$-band $Q_{\phi}$ image of HD 100453 from \cite{Benisty2017} that has been re-processed from the unscaled \texttt{FITS} image for this comparison. \textit{Right}: Polarized intensity image of the 25$^{\circ}$ model. Both the SPHERE and the model image have been scaled by $r^{2}$, with $r$ being an approximation to the deprojected distance to the star by assuming a 25$^\circ$ inclination and a flat disk. The color log-scale is arbitrary and adjusted independently in each image to show the fainter disk features. The bottom side of the disk to the south-west is clearly seen in both images.}
\end{figure}

\end{document}